\documentstyle[11pt,aaspp4]{article}

\lefthead{MIGHELL}
\righthead{K.~MIGHELL: WFPC2 OBSERVATIONS OF THE CARINA DSPH GALAXY}
\slugcomment{Accepted by the Astronomical Journal on 7 July 1997}

\begin{document}

\def\et{\hbox{\em{et~al.\ }}}
\def\eg{\hbox{e.g.\ }}
\def\ie{\hbox{i.e.\ }}
\def\vs{\hbox{vs }}
\def\bmv{\hbox{$B\!-\!V$}}
\def\ebmv{\hbox{E$(\bmv)$}}
\def\vmi{\hbox{$V\!\!-\!I$}}
\def\vmr{\hbox{$V\!\!-\!R$}}
\def\evmi{\hbox{E$(\vmi)$}}
\def\feh{\hbox{[Fe/H]}}
\def\measurement#1#2{\mbox{$[#1\small{(#2)}]$}}
\def\mvto{\hbox{$M_V^{\rm TO}$}}
\newcommand{\myfirstuse}{}
\def\mfu#1{}
\def\mb#1{#1}

\title{
WFPC2 OBSERVATIONS OF THE CARINA\\
DWARF SPHEROIDAL GALAXY\altaffilmark{1}
}

\author{
\sc
Kenneth J. Mighell\altaffilmark{2}
}
\affil{
\small
Kitt Peak National Observatory\altaffilmark{3}\mb{,}
National Optical Astronomy Observatories\mb{,}
P.O. Box 26732, Tucson, AZ~~85726-6732\mb{;}
Electronic mail: mighell$@$noao.edu
}

\altaffiltext{1}{
Based on observations made with the NASA/ESA
{\em{Hubble Space Telescope}},
obtained from the data archive at the Space Telescope
Science Institute,
which is operated by the Association of
Universities for Research in Astronomy, Inc.\ under NASA
contract NAS5-26555.
}
\altaffiltext{2}{
Guest User, Canadian Astronomy Data Center, which is operated by the
Dominion Astrophysical Observatory for the National Research Council of
Canada's Herzberg Institute of Astrophysics.
}
\altaffiltext{3}{
Kitt Peak National Observatory,
National Optical Astronomy Observatories,
is operated by the
Association of Universities for Research in Astronomy, Inc.\ (AURA)
under cooperative agreement with the
National Science Foundation.
}

\begin{abstract}
\noindent
We present our analysis of
{\sl{Hubble Space Telescope}} Wide Field Planetary Camera 2
observations
in
F555W ($\sim$$V$)
and
F814W ($\sim$$I$)
of the Carina dwarf spheroidal galaxy.
The resulting $V$ vs $\vmi$ color-magnitude diagrams
reach $V\!\approx\!27.1$ mag.
The reddening of Carina is estimated to be
$\evmi=0.08\!\pm\!0.02$ mag.
A new estimate of the distance modulus of Carina,
$(m-M)_0^{~} = 19.87\!\pm\!0.11$ mag,
has been derived primarily from existing photometry in the literature.
The apparent distance moduli in $V$ and $I$ were determined to be
$(m-M)_V^{~} = 20.05\!\pm\!0.11$ mag
and
$(m-M)_I^{~} = 19.98\!\pm\!0.12$ mag,
respectively.
These determinations assumed that Carina has a metallicity of
$\feh=-1.9\!\pm\!0.2$ dex.
This space-based observation,
when combined with previous ground-based observations,
is consistent with
(but does not necessarily prove)
the following star formation scenario.
The Carina dwarf spheroidal galaxy formed its old stellar population in a
short burst ($\lesssim$3 Gyr) at about the same time the Milky Way formed
its globular clusters.
The dominant burst of intermediate-age star formation then began
in the central region of the galaxy where stars formed for
several billion years before the process of star formation became
efficient enough in the outer regions of the galaxy to allow for the
formation of large numbers of stars.
There has been negligible star formation during the last few billion years.
This observation provides evidence that
at least some dwarf galaxies can have complex
global star formation histories with local
variations of the rate of star formation as a function of time and
position within the galaxy.
\end{abstract}

\newpage
\section{INTRODUCTION}
Many of the Galactic dwarf spheroidal satellites have had complicated
star formation histories
with periods of star formation lasting many billions of years
(e.g.\
Carina:
Mighell \cite{mi1990a},
Mighell \& Butcher \cite{mibu1992},
Smecker-Hane \et \cite{smet1994}, \cite{smet1996};
Fornax:
Beauchamp \et \cite{beet1995};
Leo I:
Lee \et \cite{leet1993a};
Leo II:
Mighell \& Rich \cite{miri1996}).
These galaxies have a range of ages and/or metallicities despite their
low stellar densities and escape velocities.  Our current understanding of
star formation suggests that the stars in dwarf spheroidal galaxies
could not have formed at the currently observed low stellar densities.
Faber \& Lin (\cite{fali1983})
suggested that the presence of a significant amount of dark matter is the
key feature which distinguishes dwarf spheroidal galaxies from star
clusters.
Mateo \et (\cite{maet1993}) and
Vogt \et (\cite{voet1995}) suggested that
all small galaxies have similar and large dark-matter components.
The Local Group dwarf spheroidal galaxies present a challenge to
our current understanding of the formation and evolution of dwarf galaxies.

Precision photometry of crowded stellar fields seen in
Local Group galaxies is very difficult even with the best ground-based
new-technology telescopes.
The high angular resolution of the
{\sl{Hubble Space Telescope}} Wide Field Planetary Camera 2 (WFPC2)
instrument provides an excellent tool for the analysis of stellar populations
in nearby Local Group galaxies.
The WFPC2 instrument is now being used to make
deep color-magnitude diagrams of a wide variety of stellar populations
in Local Group galaxies.
These deep color-magnitude diagrams are required for us to
fully understand the star formation history and evolution
of our neighboring galaxies in the Local Group.

In this work we extend our understanding of the complex star formation
history of the Carina dwarf spheroidal galaxy
using archival WFPC2 data.
Section 2 is a discussion of the observations and photometric reductions.
We review previous work Sec.\ 3 and present our results in Sec.\ 4.
We summarize the results of the paper in Sec.\ 5.
Appendix A describes how we revised
the distance modulus for Carina primarily from
existing data in the literature.

\section{OBSERVATIONS AND DATA REDUCTION}

The Carina dwarf spheroidal galaxy
was observed with the
{\sl{Hubble Space Telescope}} Wide Field Planetary Camera 2 (WFPC2)
on 1995 January 3 through the
F555W ($\sim$$V$)
and
F814W ($\sim$$I$)
filters.
The WFPC2 WFALL aperture
(Biretta \et \cite{biet1996})
was centered on the target position of
$\alpha = 06^{\rm h}\ 41^{\rm m}\ 49^{\rm s}$
and
$\delta = -50\arcdeg\ 58\arcmin\ 11\arcsec$
(J2000.0)
and
four low-gain observations were obtained:
two 1100 s exposures in F814W (datasets U2LB0102T and U2LB0103T)
and
two 1100 s exposures in F555W (datasets: U2LB0105T and U2LB0106T).
These observations were secured as
part of the {\sl HST} Cycle 4 program GTO/WFC 5637
and were placed in the public data archive
at the Space Telescope Science Institute on 1996 January 3.
The datasets were recalibrated at the Canadian Astronomy Data Centre
on 1996 August 13 and were retrieved electronically that same day.
The cosmic rays were
removed by using the {\bf{IRAF}}\footnote{
IRAF is distributed by the National Optical Astronomy
Observatories, which is operated by the Association of Universities for
Research in Astronomy, Inc.\ under cooperative agreement with the
National Science Foundation.}{\bf{/STSDAS}} task {\bf{crrej}} to
make one clean F814W observation of 2200 s and one clean F555W
observation of 2200 s.
We only present the analysis of data obtained from the WF cameras
in this paper.

Unsharp mask images of the clean F555W and F814W observations were made
using the LPD (low-pass difference) digital filter
which was designed by Mighell
to optimize the detection of faint stars in {\sl HST} WF/PC and WFPC2
images (Appendix A of Mighell \& Rich \cite{miri1995}, and references therein).
The two unsharp mask images were then added together to create a
master unsharp mask image of each WF CCD.
A simple peak detector algorithm was then used on the master unsharp
images to create a
list of point source candidates
with coordinates $60 \leq x \leq 790$ and $60 \leq y \leq 790$
on each WF CCD.
This allowed the use of almost the entire field-of-view of each WF camera
while avoiding edge-effects in the outer regions.

The data were analyzed with
the {\bf{CCDCAP}} digital circular aperture
photometry code developed by Mighell to analyze
{\sl{HST}} WF/PC and WFPC2
observations (Mighell \et \cite{miet1996}, and references therein).
A fixed aperture with a radius of 1.8 pixels was used for all stars
on the WF CCDs.
The local background level was determined from a robust estimate
of the mean intensity value of all pixels between 1.8 and 6.0 pixels
from the center of the circular
stellar aperture.
Point source candidates were rejected if either
one of two criteria was
satisfied:
(1) the measured signal-to-noise ratio of
either instrumental magnitude was less than 5;
or (2) the center of the aperture [which was allowed to move in order to
maximize the signal-to-noise ratio ($SNR$) ] changed by more than 1 pixel
from its detected position on the unsharp mask.
Almost all photon noise spikes
(due mainly to the background sky and diffraction spikes)
were automatically eliminated by using these criteria.

Observed WFPC2 point spread functions (PSFs)
vary significantly
with wavelength, field position, and time (Holtzman \et \cite{hoet1995a}).
There were not enough bright isolated stars in these WFPC2 observations
to adequately measure the variation of the point spread function
across each WF CCD using the observations themselves.
We measured artificial point spread functions synthesized by the
{\bf{Tiny Tim Version 4.0b}}
software package (Krist \cite{kr1993}, \cite{kr1994})
to determine the aperture corrections,
$\Delta_r$,
required to convert instrumental magnitudes
measured with an aperture of radius 1.8 pixels (0.18\arcsec)
to a standard aperture of radius 5.0 pixels (1\arcsec\ diameter).
A catalog of 289 synthetic M-giant point spread functions was created with
a $17 \times 17$ square grid
for each
filter (F814W and F555W)
and CCD (WF2, WF3, and WF4).
The spatial resolution of one synthetic PSF
every 50 pixels in $x$ and $y$ allowed for the determination
of aperture corrections
for any star in the entire WFPC2 field-of-view to have a
spatial resolution of $\lesssim$35 pixels.
The aperture corrections, by definition, are always negative.
The average aperture corrections,
$\langle \Delta_r \rangle$,
are listed in Table \ref{tbl-1}
{\mfu{Tab\ref{tbl-1}}.

\placetable{tbl-1}

The WFPC2 point spread functions can vary with time due to
spacecraft jitter during exposures and small focus
changes caused by the {\sl{HST}} expanding and contracting (``breathing'')
once every orbit.
These temporal variations of WFPC2 PSFs
can cause small, but significant, systematic
offsets in the photometric zeropoints when small apertures are used.
Fortunately,
these systematic offsets can be easily calibrated away by
simply measuring
bright isolated stars on each CCD twice: once with the small aperture
and again with a larger aperture.
The robust mean magnitude difference
between the large and small apertures
is then the
zero-order aperture correction,
$\delta_r$,
for the small aperture which, by definition,
can be positive or negative.
Zero-order aperture corrections
are generally small for long exposures,
however, they
can be large for short exposures that were obtained while the WFPC2 was
slightly out of focus (by a few microns)
due to the expansion/contraction of the {\sl HST} during
its normal breathing cycle.
The zero-order aperture corrections for these observations
($\delta_r$ : see Table \ref{tbl-1})
were computed using a large aperture with a radius of 5.0 pixels (0.50\arcsec)
and a background annulus of
$5.0\leq r_{\rm sky}\leq8.0$ pixels.

The Charge Transfer Effect was
removed from the instrumental magnitudes
by using a 4\% uniform wedge
along the Y-axis of each WF chip as described in
Holtzman \et (\cite{hoet1995b}).

We used the standard WFPC2 color system
(Holtzman \et \cite{hoet1995b})
which is defined using apertures 1\arcsec\ in diameter
containing about 90\% of the total flux from a star.
The instrumental magnitudes,
$v_r$ and $i_r$,
were transformed to Johnson $V$ and Cousins $I$ magnitudes
using the following equations
\begin{eqnarray}
V
&=
&v_r + \Delta_r + \delta_r \nonumber\\
&&
+~{-0.052 \brack \pm0.007}(\vmi)
+~{ 0.027 \brack \pm0.002}(\vmi)^2
+~{21.725 \brack \pm0.005 }
\end{eqnarray}
and
\begin{eqnarray}
I
& =
&i_r + \Delta_r + \delta_r \nonumber\\
&&
+~{-0.062 \brack \pm0.009}(\vmi)
+~{ 0.025 \brack \pm0.002}(\vmi)^2
+~{20.839 \brack \pm0.006 }
\end{eqnarray}
where an instrumental magnitude of zero is defined as one DN/sec at
the high gain state ($\sim$14 e$^-$/DN).
The constants in brackets come from Table 7 of Holtzman \et (\cite{hoet1995b}).

We present our stellar photometry of 3302 stars in the central region of the
Carina dwarf spheroidal galaxy with
SNR$\,\geq\,$5 in both the F555W and F814W filters
in the CD--ROM Table 1 found in the
AAS CD-ROM Series,
Vol.\ {\framebox{XX}}, 199{\framebox{X}}
{\mfu{CDTab1}}.
A printed version showing the first 20 stars is given in
Table \ref{tbl-2}{\mfu{Tab\ref{tbl-2}}.
The first column gives the identification (ID) of the star.
The left-most digit of the ID gives the WFPC2 chip number (2, 3, or 4)
where the star was found.
The right-most 4 digits gives the $x$ coordinate of the star multiplied
by 10.  The remaining 4 digits gives the $y$ coordinate of the star multiplied
by 10.  All positions are given with respect to the U2LB0102T
dataset.  For example, the first star in Table \ref{tbl-2} has an ID of
206092291 which indicates that it has the $(x,y)$ position of $(229.1,60.9)$
on the WF2 CCD of the U2LB0102T dataset.  The second and third columns
give the $V$ magnitude and its r.m.s.\ photometric error $\sigma_V$.
Likewise, the fourth and fifth columns give the $(\vmi)$ color and its
r.m.s. photometric error $\sigma_{(V\!-I)}$.

\placetable{tbl-2}


\section{PREVIOUS STUDIES}

The Carina dwarf spheroidal galaxy
(Cannon \et \cite{caet1977})
is one of the Galaxy's nine
dwarf spheroidal companions\footnote{
In order of discovery,
the nine dSph companions of the Milky Way are
Sculptor,
Fornax,
Leo I,
Leo II,
Ursa Minor,
Draco,
Carina,
Sextans,
and
Sagittarius.
}, and it was the
first of these in which an extended star formation history was
conclusively detected.

Mould \& Aaronson (\cite{moaa1983})
obtained a $V$ vs $B-V$ color-magnitude diagram (CMD)
of the
Carina dwarf spheroidal galaxy that
reached $V$$\approx\,$$24$ mag and
showed a main-sequence turnoff at $V_{\rm{TO}} = 23.0\!\pm\!0.2$ mag.
Carina was found to be metal-poor with a metallicity of
$\feh$$=$$\,-1.9\pm0.2$ dex.
A stubby, red horizontal branch (HB) was found at $V_{\rm{HB}}=20.5$ mag
and a distance modulus of
$(m-M)_0=19.8$ mag was derived
by assuming an absolute $V$ magnitude for the horizontal branch of
$M_V^{\rm{HB}}=0.6$ mag.
Isochrone fitting and a horizontal-branch to
main-sequence turnoff difference of
$\Delta V_{\rm TO}^{\rm HB}\approx2.5$ mag yielded
an age estimate of $7.5\!\pm\!1.5$ Gyr for the bulk of the stellar population
of Carina.

Deep $VR$ photometry ($V$$\leq\,$$25.0$ mag) of
Carina
by Mighell (\cite{mi1990a})
allowed him to produce
a color-magnitude diagram
that reached 2 mag below the main-sequence turnoff of the intermediate-age
stellar population.  This CMD
showed the existence of the  old stellar population suggested by
the presence of RR Lyraes in the galaxy
(Saha \et \cite{saet1986}).
Mighell estimated that the main-sequence turnoff magnitude of the
old stellar population is $23.5\!\leq\!V_{\rm{TO}}\!\leq\!23.8$ mag.
While the stellar population of Carina
is dominated by intermediate-age (6--9 Gyr) stars,
Mighell
found that
17$\pm$4 percent of the
total stellar population is older than 13 Gyr.
{}From the analysis of the distribution of colors near the main-sequence
turnoff, Mighell
suggested that
the old population in Carina was probably formed in a single burst lasting
less than 3 Gyr and that the rate of star formation became negligibly small
until the dominant intermediate-age population formed 6--9 Gyr ago.

Mighell (\cite{mi1990b}) and Mighell \& Butcher (\cite{mibu1992})
compared the deep $V$ stellar luminosity function of Carina
(Mighell \cite{mi1990c})
with theoretical stellar luminosity functions from
{\it{The Revised Yale Isochrones and Luminosity Functions}}
(Green \et \cite{gret1987}, hearafter RYI)
and a star formation history of Carina was derived that was
fully consistent with star formation histories previously
determined by analyzing color-magnitude diagrams.

Smecker-Hane \et (\cite{smet1994})
surveyed a large fraction of Carina
and produced a shallow color-magnitude diagram
($V$$\leq$$\,23.0$ mag)
which shows a
second, morphologically distinct, horizontal branch at
$V_{\rm{HB}}^{~}=20.65\!\pm\!0.05$ mag.
The fainter horizontal branch
has blue and red HB stars with an RR Lyrae strip;
a morphology that is typical of old, metal-poor globular clusters.
A distance modulus of
$(m-M)_0=20.09\!\pm\!0.06$ for Carina
was derived from the new determination of the apparent $V$ magnitude
of the horizontal branch
and the $I$ magnitude of the tip of the red giant branch
($I^{~}_{\rm{TRGB}} = 16.15\!\pm\!0.05$ mag).
Smecker-Hane \et (\cite{smet1996}) recently presented
a preliminary deep $R$ vs $B\!-\!R$ color-magnitude diagram
($R$$\leq$$\,25.0$ mag) of Carina
which clearly shows the intermediate-age and old stellar population
turnoffs.
This preliminary CMD also shows the presence of a small population of stars
possibly as young as 2 Gyr old.

The simplest explanation of Carina's complex horizontal branch
morphology is that the stubby, red HB (at $V\!=\!20.5$) is associated with its
dominant intermediate-age stellar population and its
fainter HB (at $V\!=\!20.65\!\pm\!0.05$)
is associated with its old stellar population.
Mighell (\cite{mi1990a})
showed that most of the stars in Carina
belong to its intermediate-age (6--9 Gyr) stellar population
and
Smecker-Hane \et (\cite{smet1994})
report that $\sim$72\% of the HB stars
are associated with the stubby, red HB.
Theoretical horizontal branch models
predict that the absolute $V$ magnitude
of the horizontal branch increases (fades) with age.
Sarajedini \et (\cite{saet1995}),
using the theoretical HB models of
Lee \et (\cite{leet1990}, \cite{leet1994}),
showed that the $M_V^{\rm{HB}}$ of a $\sim$15 Gyr
metal-poor ($\feh=-1.8$) stellar population is
$\sim$0.2 mag greater (fainter) than the
$M_V^{\rm{HB}}$ of a $\sim$7 Gyr stellar population
with the same metallicity.
This theoretical difference in the absolute $V$ magnitude
($\Delta M_V^{\rm{HB}}\approx0.2$ mag)
is quite close to the observed $V$ magnitude difference found
between Carina's two horizontal branches
($\Delta V_{\rm{HB}}^{~}\approx0.15\!\pm\!0.05$ mag).
If the faint HB is associated with Carina's small old stellar population,
then it is really not surprising that
Mould \& Aaronson (\cite{moaa1983})
and
Mighell (\cite{mi1990a})
did not detect the faint HB ---
the small field-of-view of their CCD observations
prevented them from observing
a statistically significant number of these rare stars.

Mighell's (1990a) estimate that 17$\pm$4 percent of the
total stellar population is older than 13 Gyr should probably
be interpreted as a {\em lower limit} of the size of Carina's
old stellar population since abundant intermediate-age stars
near the main-sequence turnoff
were compared with relatively rarer old subgiant branch stars.
Mighell \& Butcher (\cite{mibu1992}) found that
the maximum size of Carina's old stellar population
is, conservatively, $40\!\pm\!5$\% of the total stellar population.
Smecker-Hane \et (\cite{smet1994}) found that the mean fraction
of old horizontal branch stars in Carina
is $27\!\pm\!7$\% with no systematic radial dependence within 10 arcmin
of the center of the galaxy.
This value should probably be taken as an {\em upper limit} of the
size of Carina's old stellar population because
the old HB stars live longer as horizontal branch stars
than do the more massive intermediate-age HB stars.
In support of these lower and upper limits, we note
that the preliminary analysis of
Smecker-Hane (\cite{sm1997})
suggests that the old stellar population of
Carina accounts for 20\% (no error is given) of the stellar population.

Da Costa and Hatzidimitriou have obtained AAT spectra at the
\ion{Ca}{2} triplet
(near 8600~\AA: Armandroff \& Da Costa \cite{ardaco1991})
of a sample of candidate Carina members (Da Costa \cite{daco1993}).
Radial velocity measurements from the spectra confirmed that 15 stars
were members of the Carina dwarf spheroidal galaxy.
The mean abundance of the 15 member stars was
$\feh = -1.88\!\pm\!0.08$ dex.
There was one star,
with $\feh = -2.32\!\pm\!0.10$,
which is substantially more metal-poor than the other 14 stars in the sample.
The intrinsic abundance spread seen in the other 14 stars in the sample
is small: $\sigma_{\rm{[Fe/H]}}=0.14$ dex.
This spectroscopic metallicity determination
agrees well with previous estimates based on color and magnitude differences
derived from CCD stellar photometry of the red giant branch,
horizontal branch and the main-sequence turnoff region of Carina
($\feh = -1.9\!\pm\!0.2$ : Mould \& Aaronson \cite{moaa1983};
$\feh = -1.75\!\pm\!0.2$ : Mighell \cite{mi1990a};
$\feh = -2.0$ : Smecker-Hane \et \cite{smet1994}).
For the rest of this paper, we will assume that the metallicity of
the Carina dwarf spheroidal galaxy is $\feh = -1.9\!\pm\!0.2$ dex.

\section{THE CENTRAL REGION OF CARINA}

The $V$ vs $\vmi$ color-magnitude diagram and
apparent $V$ stellar luminosity
function of the observed stellar
field in the Carina dwarf spheroidal galaxy
are displayed in
Fig.\ \ref{fig-1}{\mfu{Fig\ref{fig-1}}}.
This figure includes over 3,185 stars down to a limiting
magnitude of $V\!\approx\!27.1$ mag on
the main sequence of Carina.
The galactic latitude of Carina is low
($l\!=\!260.1\arcdeg\!\!, b\!=\!22.2\arcdeg$)
and contamination by field stars can be substantial
(e.g.
Fig.\ 3 of Mould \& Aaronson \cite{moaa1983};
Fig.\ 2 of Smecker-Hane \et \cite{smet1994}).
The red giant branch in Fig.\ \ref{fig-1}
is sparse and there is probably some contamination by field stars
even though the field is located near the center of Carina
[$r \approx 0.1 r_c$\footnote{
The core and tidal radii of Carina are
$r_c=8.8\!\pm\!1.2$
arcmin
and
$r_t=28.8\!\pm\!3.6$
arcmin, respectively (Irwin \& Hatzidimitriou \cite{irha1995}).}].
The clump of stars seen at $V\!\approx\!20.5$ mag and
$(\vmi)\!\approx\!0.9$ mag
are helium core-burning stars belonging to
the intermediate-age stellar population.
The field-of-view of this observation (4.44 arcmin$^2$)
is 3.79 times smaller than the field-of-view of Mighell (\cite{mi1990a})
who found few, if any, horizontal branch stars belonging to the old
stellar population of Carina.
We would thus expect to find a statistically insignificant
number of horizontal branch stars associated with the old stellar
population of Carina in these WFPC2 observations.

\placefigure{fig-1}

Many stars of varying ages are seen in Fig.\ \ref{fig-1} to
have evolved off the main-sequence.
The width of the main-sequence turnoff region cannot be entirely due to
photometric scatter because the turnoff region is significantly wider than the
photometric errors at those magnitudes.
Main sequences can be broadened by
large spreads in metallicity,
large variations in reddening,
or
long periods of star formation.
Smecker-Hane \et (\cite{smet1996}) have estimated that the metallicity
spread in Carina is small ($\lesssim\!0.2$ dex)
and we show in Appendix A that
the reddening of central region of Carina is low
[ $\evmi\!=\!0.08\!\pm\!0.02$ mag].
It is therefore unlikely that the
observed width of the Carina main sequence is due to a metallicity spread
or variations in reddening.
Carina has been forming stars for many billions of years
(Mighell 1990a), and thus
the simplest explanation for Carina's wide main sequence
is that it is due to a long period of star formation.

Previous deep color-magnitude diagrams of Carina
show a gap in the subgiant branch region redward of the
main-sequence turnoff of the intermediate-age stellar population
(Fig.\ 9 of Mighell \cite{mi1990a},
Fig.\ 1 of Smecker-Hane \et \cite{smet1996}).
Mighell (\cite{mi1990a})
argued that this gap was due to negligible star formation
between the bursts that created the old and intermediate-age
populations.
The gap seen in the deep color-magnitude diagram of
Smecker-Hane \et (\cite{smet1996}) is not completely empty.  This
suggests that star formation might never have actually stopped between the
two major bursts that created the old and intermediate-age stellar
populations.  However, some of the gap stars are undoubtably
foreground stars in the Milky Way.
It is difficult to accurately determine the amount of Galactic field star
contamination in a large survey of Carina
without deep observations (hopefully of equal or greater areal coverage)
of nearby fields well beyond the tidal radius of
the galaxy.

The apparent $V$ stellar luminosity function (SLF)
of this WFPC2 observation of the central region of the Carina
dwarf spheroidal galaxy is redisplayed
in greater detail in Fig.\ \ref{fig-2}{\mfu{Fig\ref{fig-2}}}.
The shape of the faint end of the color-magnitude diagram in
Fig.\ \ref{fig-1} is determined mainly by
the requirement that the signal-to-noise ratio of each star
must be at least 5 in both the F555W and F814W filters.
The interpretive value of the apparent $V$ SLF
at magnitudes fainter than $V\!=\!26.0$ mag will be limited
by these selection effects.
The open circles show the apparent $V$ SLF for stars in the Fig.\ \ref{fig-1}
CMD with the additional constraint that SNR$\,\geq\!10$ in the F555W filter.
The low-SNR and the high-SNR
apparent $V$ SLFs are identical in the magnitude range
$19.8\!\leq\!V\!\leq\!26.0$ with the statistically insignificant
difference in the $V\!=\!25.7$ mag bin where the low-SNR SLF
has one more star.

\placefigure{fig-2}

Figure \ref{fig-2} shows that the published completeness-corrected
$V$ stellar luminosity function of Mighell (\cite{mi1990c}) is well matched
by the apparent $V$
stellar luminosity function of the WFPC2 field when
it is scaled down by a factor of 3.2 to give
the same number of stars in the $V$ magnitude range
$22.75\!\leq\!V\!\leq\!23.25$ which is
the main-sequence turnoff region of the
intermediate-age stellar population.
The 6, 8, and 14 Gyr theoretical RYI
stellar luminosity functions for
$Z=0.00025$ ($\feh = -1.9$ dex) and $Y=0.24$
are displayed in Fig.\ \ref{fig-2} for comparative purposes.
The apparent $V$ WFPC2 Carina SLF
appears to be well bracketed by the 6 and 8 Gyr
RYI stellar luminosity functions in the
magnitude range $22.5\leq V \leq 25.0$ mag.
This suggests that the {\em{median}} age of
the {\em intermediate-age} stellar population in the central region
of Carina is about $7\!\pm\!1$ Gyr on the RYI age scale.

Recent theoretical isochrones generally
predict ages for Population II stars
that are about 7--8\% younger than the ages predicted by
theoretical isochrones developed in the 1980's
(private communications:
VandenBerg 1996,
Demarque 1996,
Chaboyer 1997).
The younger ages are due principally to improvements of the input physics
(\eg the equation of state).
A better estimate of the median age of the intermediate-age stellar
population in the central region of Carina
would thus be $\sim\!6.5\!\pm\!0.9$ Gyr.

The $V$ vs $\vmi$ color-magnitude diagram of the main-sequence turnoff
region of the Carina dwarf spheroidal galaxy
is displayed in Fig.\ \ref{fig-3}{\mfu{Fig\ref{fig-3}}}\ with
the theoretical RYI isochrones for
the metallicity $Z=0.00025$ ($\feh=-1.9$ dex)
and helium abundance $Y=0.24$.
This figure suggests that
the bulk of star formation in the central region of Carina
ended about 4 Gyr ago.

\placefigure{fig-3}

The main-sequence turnoff region
is the most sensitive area of the color-magnitude diagram
for the determination of the absolute age of stellar populations.
Provided that the metallicity, apparent distance modulus
and reddening of a star is known,
one can determine the age of a subgiant branch star, in principle,
by finding two isochrones that constrain the
star in magnitude and color within the subgiant branch region.
The age of the star can be estimated
as being the average of the ages of the two isochrones with the estimate of the
r.m.s. error being one half the difference in the ages of the two isochrones.
Measurement (\eg photometric) errors and systematic errors
(\eg
uncertainties associated with the apparent distance modulus, reddening, and
the physics and approximations of the stellar evolution model)
will, of course, increase the uncertainty associated with the derived ages.
This method of determining stellar ages can also be
used to derive the star formation histories of simple stellar populations
given high signal-to-noise stellar observations of systems
that have experienced little or no chemical evolution.

The ages of subgiant branch stars
in the central region of Carina
were estimated in the following manner.
The $V$ vs $\vmi$ color-magnitude diagram of the main-sequence turnoff
region of the Carina dwarf spheroidal galaxy
is displayed in Fig.\ \ref{fig-4}{\mfu{Fig\ref{fig-4}}}\ with
the theoretical RYI isochrones for
the metallicity $Z=0.00025$ ($\feh=-1.9$ dex)
and helium abundance $Y=0.24$ at ages from 2 to 18 Gyrs in 1 Gyr increments
(heavy solid curves, top to bottom).
Isochrones were derived at 0.25 Gyr intervals from 2 to 18 Gyr
by interpolating the standard RYI isochrones
for $Z=0.00025$ and $Y=0.24$ at
ages $2, 3, 4, \ldots, 18$ Gyr.
These interpolated isochrones are
displayed in Fig.\ \ref{fig-4} (light curves, top to bottom)
from a point on the isochrone near the main-sequence turnoff\footnote{
Defined as the location on the isochrone
where the RYI equivalent evolutionary phase is EEP$= 67$.}
to the base of the red giant branch
which was defined as the place on the isochrone where the dereddened color
$(\vmi)_0$ is $0.70$ mag.
The initial age estimate,
$t_{\rm{Gyr}}^\prime\,$,
of each of the 152 stars within the bounded region of Fig.\ \ref{fig-4}
is given in Table \ref{tbl-3}{\mfu{Tab\ref{tbl-3}}}.

\placefigure{fig-4}

\placetable{tbl-3}

Better age estimates can be derived from Monte Carlo simulations
using the photometric data given in CD-ROM Table 1.
One thousand ``observations'' of each of the 152 stars in
the bounded region of
Fig.\ \ref{fig-4}
were simulated (see Fig.\ \ref{fig-5}{\mfu{Fig\ref{fig-5}})
by using two-dimensional Gaussian distributions with
mean $V$ magnitudes and $\vmi$ colors
and standard deviations in $V$ and $\vmi$
defined as the respective
measured values of $V$, $\vmi$, $\sigma_V$, $\sigma_{(V\!-I)}$,
which were given originally in CD-ROM Table 1 and duplicated in
Table \ref{tbl-3}.
The mean stellar age (in Gyr on the RYI age scale)
and r.m.s. error for each of the 152 stars in the
bounded region is given in Table \ref{tbl-3} columns
$\langle t_{\rm{Gyr}}\!\rangle$
and
$\sigma_{\langle t_{\rm{Gyr}}\!\rangle}$,
respectively.
The mean stellar ages were determined by analyzing only the simulated
observations that were found to be inside the bounded region.

\placefigure{fig-5}

The r.m.s.\ errors of the stellar age estimates increase with age
(see Fig.\ \ref{fig-6}{\mfu{Fig\ref{fig-6}}})
because the spacing between theoretical isochrones generally decreases with
increasing age and photometric errors increase with
apparent magnitude.
The brightest subgiant branch star
in the bounded region of Fig.\ \ref{fig-5} is star
256051426 which has an age estimate of $3.13\!\pm\!0.14$ Gyr.
The small error estimate for this star
is due to the fact that all of the 1000
simulated ``observations'' were found in the fifth age bin (5.00--5.25 Gyr).
Star 317853886 is $1.289\!\pm\!0.015$ mag fainter
and its 1000 simulated observations were spread across 4 bins which
translates to an r.m.s.\ error of 0.29 Gyr for a
mean stellar age of 9.54 Gyr on the RYI age scale.
Stars fainter than $V\!\approx\!23$ mag
will increasingly suffer from ``bin hopping'' due primarily to color errors.

\placefigure{fig-6}

The mean estimated ages (in Gyr on the RYI age scale) and r.m.s.\ errors
of all 152 stars in the bounded region of Fig.\ 4
are shown in Fig.\ \ref{fig-7}{\mfu{Fig\ref{fig-7}}}.
The r.m.s.\ errors increase with age, as expected, until the mean
estimated age of $\sim$$15$ Gyr.
which occurs near the bottom of the bounded region.
This region in Fig.\ \ref{fig-5} shows that many of the
simulated observations are found outside of the bounded region,
which suggests that mean estimated ages greater than $\sim$$15$
Gyr are probably not reliable due to incompleteness effects.

\placefigure{fig-7}

How many stars in the bounded region actually belong there?  While the vast
majority of the simulated ``observations'' shown in in Fig.\ \ref{fig-5}
are found within the bounded region, some of the stars are found beyond
the edges of the bounded region.  It is quite likely that some of the stars
within the bounded region have been photometrically scattered {\em into}
the bounded region.  Unfortunately, it is impossible to precisely say which
stars have done so without {\em a priori} knowledge of the stellar magnitudes
and colors.  We introduced the above Monte Carlo simulations so that we could
obtain better stellar age estimates.  We now reuse these same simulations
in an attempt to determine just how many stars in the bounded region actually
belong there.

A total of 11,104 simulated observations out of a grand total of 152,000
were found outside of the bounded region in Fig.\ \ref{fig-5}.
This suggests that probably no more than
$\sim$92\% ($\sim$140)
of the 152 stars found within the bounded region actually belong there.
The recovery probabilities, $p$, of all 152 stars in the bounded region
of Fig.\ \ref{fig-4} is given in Table \ref{tbl-3}.
There are 77 stars in Table \ref{tbl-3}
which had all 1000 of their simulated observations found within the bounded
region ($p\!\equiv\!1$).
Only 12 stars had fewer than 700 of their 1000 simulated observations
found within the bounded region ($p\!<\!0.700$).
The recovery probabilities given in Table \ref{tbl-3} are
summarized in Table \ref{tbl-4}{\mfu{Tab\ref{tbl-4}}}.
This table shows that one star (225033399)
had a recovery probability of $p\!=\!0.897$.
Of the 116 stars with recovery probabilities $p\!\geq\!0.897$
(shown in Fig.\ \ref{fig-8}{\mfu{Fig\ref{fig-8}}})
only 1105 simulated observations (out of a grand total of 116,000)
were found outside of the bounded region.
Table \ref{tbl-4} also gives the total recovery probabilities, $p^\prime$,
which is defined as
$p^\prime
\equiv
{
\sum\!{\rm{in}}
}
/
{\left(
\sum\!{\rm{in}}
+
\sum\!{\rm{out}}
\right)}$
where
$\sum\!{\rm{in}}$
and
$\sum\!{\rm{out}}$
are given in Table \ref{tbl-4}.
Using the example above, we see that $p^\prime\!>\!0.990$ for the
the 116 stars with recovery probabilities $p\!\geq\!0.897$.
This implies that there is a 99\% probability that these 116 stars
actually belong inside the bounded region.  In other words,
only $\sim$1
star out of the 116 probably does {\em not} belong within the bounded region.

\placetable{tbl-4}

\placefigure{fig-8}

The estimated mean stellar ages
of the conservative subset of 116 subgiant branch stars
($p\!\geq\!0.897 \Rightarrow p^\prime\!\geq\!0.990$)
are displayed as the solid histogram at the bottom
of Fig.\ \ref{fig-9}{\mfu{Fig\ref{fig-9}}}.
The median age of these 116 subgiant branch stars in the central region of
Carina is
$\sim$$6.9$ Gyr
($\sim$$7.5$ Gyr on the RYI age scale).
This estimate agrees quite well with the estimate of
$\sim\!6.5\!\pm\!0.9$ Gyr for the median age of the intermediate-age
stellar population which was derived above from the analysis of the
apparent stellar luminosity function.

\placefigure{fig-9}

Any analysis of the high-resolution (0.25 Gyr bin width)
star-count histogram at the bottom of Fig.\ \ref{fig-9} will
be affected by small number (Poisson) statistics.
This analysis uncertainty can be reduced by boosting
the signal-to-noise ratio of the data at the cost of
degrading the age resolution.
Two 0.25 Gyr wide bins were added together
to produce 2 low-resolution star-count histograms with 0.5 Gyr bin widths
that are offset by one 0.25 Gyr bin from each (see Fig.\ \ref{fig-9}):
the vertical-shaded histogram
gives the counts at $2.00, 2.50, 3.00, 3.50, \ldots$ Gyr;
the horizontal-shaded histogram
gives the counts at $2.25, 2.75, 3.25, 3.75, \ldots$ Gyr.

Small number statistics preclude the derivation of a detailed
star formation history from the high-resolution star-count histogram
at the bottom of Fig.\ \ref{fig-9}.
For example, a comparison of the data between 4 and 14 Gyr with a
model which assumes 3 stars in every 0.25 Gyr bin shows that
this simple star formation history can not be excluded at the 99\%
confidence level.
The current dataset is thus fully consistent
with the statement that the central region of Carina
continuously formed stars between $\sim$3.7 and $\sim$13 Gyr ago
(4--14 Gyr on the RYI age scale).
There may well be some stars younger than $\sim$3.7 Gyr but they
are not detected in significant quantities.  And even though there
probably are stars present in this central region of Carina
that belong to the old stellar population,
little can be said about them due to their rarity and
the various interpretation
problems (e.g.\ incompleteness effects, large age uncertainties) shown above.

Mighell (\cite{mi1990a}) suggested that the Carina dwarf spheroidal
galaxy probably experienced a period of negligible stars formation
between the old and intermediate-age stellar populations.
This period was estimated to have lasted about 6--8 Gyr.
The preliminary analysis of
Smecker-Hane (\cite{sm1997}) has suggested that Carina experienced
a long quiescent period of star formation
between 7 and 10 Gyr ago.
Shifting from the LDZ distance scale (which was presumably used)
to the CDKKS distance scale
causes the distance modulus to be $\sim$$0.10$ mag smaller
at $\feh=-1.9$ dex (see Appendix A for more details).
Decreasing the distance modulus by $\sim$$0.10$ mag
increases these ages by
$\sim$9\% (\eg Eq.\ 2 of Mighell \& Butcher \cite{mibu1992})
giving an approximate age range of
7.6--10.9 Gyr
for the period of negligible star formation
between the old and intermediate-age stellar populations
of the Carina dwarf spheroidal galaxy.

Figure \ref{fig-8} shows that there are 27 stars
($\sim$23\% of 116 stars)
between estimated mean RYI
ages of 8.25 and 11.75 Gyr which is
the comparable age range on the RYI age scale.
Foreground stellar contamination
in the subgiant branch region of the Carina color-magnitude diagram
is probably quite negligible for the small field-of-view of the WFPC2
observations:
$\lesssim$0.5 foreground
stars would be expected with colors
$B\!-\!V$$<$$1.0$ mag in the relevant
magnitude range of $22.5\!\leq\!V\!\leq\!23.0$
(see Fig.\ 10 of Mighell \cite{mi1990c}).

These stars have r.m.s. errors of $\sim$0.5 Gyr ($\sim$2$\times$0.25 Gyr bins)
which could cause some younger stars to be measured as being older
because they were scattered
into the given age range.
Similarly, some older stars could have been
scattered
into the given
age range which would cause them to measured too young.
If we make the highly unlikely assumption that {\em all} of the
stars within 0.5 Gyr of both ends of the age range were due to such
scattering,
then 20 stars would still remain with estimated mean ages between
$\sim$8.1 and $\sim$10.4 Gyr
($\,8.75\!\leq\!\langle t_{\rm{Gyr}}\!\rangle\!<\!11.25$ Gyr on the RYI age
scale).
This subset of 20 stars is a significant fraction (17\%)
of the total set of 116 subgiants that have a 99\% probability of being found
within the bounded region of Fig.\ \ref{fig-4}.
The central region of Carina apparently was forming stars
during the multi-billion year ``global'' star-formation gap
between the old and intermediate-age stellar populations of Carina
which was previously found in ground-based studies.
This observation is then consistent with the interpretation that
the intermediate-age burst of star formation in Carina
began in the central region of the galaxy and proceeded there
for many billions of years before it occurred in the
outer regions of the galaxy as observed in previous ground-based studies.

This space-based observation,
when combined with previous ground-based observations,
is consistent with
(but does not necessarily prove)
the following star formation scenario.
The Carina dwarf spheroidal galaxy formed its old stellar population in a
short burst ($\lesssim$3 Gyr) at about the same time the Milky Way formed
its globular clusters.
The dominant burst of intermediate-age star formation then began
in the central region of the galaxy where stars formed for
several billion years before the process of star formation became
efficient enough in the outer regions of the galaxy to allow for the
formation of large numbers of stars.
Ground-based color-magnitude diagrams clearly show a gap in the subgiant
branch region between the old and the dominant intermediate-age stellar
populations.
The very existence of this gap allowed Mighell (\cite{mi1990a}) to
make the first estimate of the size and age of Carina's old stellar
population.
If old intermediate-age ($\,\sim\,$8--11 Gyr old) stars {\em had} formed in the
outer regions {\em in significant numbers}, then the gap would not exist
because the subgiant branch region of the
color-magnitude diagram would be filled with evolved stars.
However, the gap {\em does} exist
and that strongly suggests that the old
intermediate-age stars preferentially formed in the central region
of the galaxy.

Analysis of WFPC2 observations of the And I dwarf spheroidal galaxy
(a companion galaxy of M31)
by Da Costa \et (\cite{dacoet1996})
has shown that it too has undergone
an extended epoch of star formation.
Additionally, evidence was found that suggested that
star formation in And I was more centrally
concentrated after the initial burst of star formation.
The general star formation history of the And I dwarf spheroidal galaxy
appears to be similar to that of Leo II (Mighell \& Rich \cite{miri1996}).
Once the initial global burst of star formation in And I had ended,
it appears that the process of star formation never again
became efficient enough in the outer regions of that galaxy
to allow the formation of large numbers of stars in a secondary global
burst of star formation like that observed in the Carina dwarf.

\section{SUMMARY}

\noindent
The findings of this paper can be summarized as follows:
\begin{itemize}
\item
We have determined that the distance modulus of the Carina dwarf spheroidal
galaxy is $(m-M)_0^{~} = 19.87\!\pm\!0.11$ mag.
The apparent distance moduli in $V$ and $I$ are
$(m-M)_V^{~} = 20.05\!\pm\!0.11$ mag
and
$(m-M)_I^{~} = 19.98\!\pm\!0.12$ mag,
respectively.
The reddening of Carina is estimated to be
$\evmi=0.08\!\pm\!0.02$ mag.
These determinations
assumed that metallicity of the galaxy is $\feh=-1.9\!\pm\!0.2$ dex.
\item
The central region of Carina has apparently been forming stars from at
least $\sim$10.4 Gyr ago to $\sim$3.7 Gyr ago;
the data is fully consistent with the interpretation
of continuous star formation in this region since at least $\sim$13 Gyr ago.
The median age of the intermediate-age stellar
population in the central region of Carina
is $\sim\!6.5\!\pm\!0.9$ Gyr.
Some stars younger than $\sim$3.7 Gyr may well exist in the central region
but they are not detected in significant quantities.
Little can be determined about the old stellar population due mainly to
small field-of-view of the current observation and the
large photometric uncertainties observed at the old turnoff
magnitudes.
\item
This space-based observation,
when combined with ground-based results,
is consistent with the interpretation that
the dominant intermediate-age burst of star formation in Carina
began in the central region of the galaxy
where stars formed for
several billion years before the process of star formation became
efficient enough in the outer regions of the galaxy to allow for the
formation of large numbers of intermediate-age stars.
\end{itemize}

Our understanding of the star formation history
of Carina has been driven by technology.
The observations of
Mould \& Aaronson (\cite{moaa1983}),
Mighell (\cite{mi1990a}),
and
Smecker-Hane \et (\cite{smet1996}) used successive generations
of CCD camera systems on 4-m class telescopes.
The high angular resolution of the the {\sl HST} WFPC2 instrument
has provided the next jump in photometric accuracy.
Each step along the way, better observations have allowed
researchers to study the star formation history of Carina
in greater and greater detail.
We have studied the star formation
history of the central region of the Carina dwarf spheroidal galaxy
from the analysis of only 116 stars in a very small field-of-view
which surveyed less than 0.5\% of Carina.
One must be cautious not to over interpret the star formation history of
a small fraction of a galaxy as being the global star formation history
of that galaxy.
Small but significant populations may be missed.  For example,
Smecker-Hane \et (\cite{smet1996}) claim to have detected a
2 Gyr stellar population ($\sim$3 Gyr on the CDKKS distance scale).
We can neither confirm or reject this finding based on
the small field-of-view of these WFPC2 observations.

The analysis technique presented in this paper has the potential of
being able to produce a detailed description of the
the complex global star formation history of Carina
with a (theoretical) age resolution much better than 1 Gyr.
The present effort was hampered by having only 116 subgiants to study.
One can get more stars by surveying more of the galaxy or
by obtaining better photometry through longer exposures.
A larger and deeper space-based survey would find more subgiants
and the longer exposures would
enable more stars to be retained at
a 99\% confidence level due to smaller photometric scatter.
Larger and more accurate surveys of Carina
will require the use of state-of-the-art theoretical
stellar evolution models.
Most of the improvements of the new stellar evolutionary models
has occurred in the understanding of stars that are more metal rich
than the stellar populations of Carina.
The newer isochrones generally produce younger ages
for metal-poor globular cluster stars, however the
shape of the isochrones of old low-mass Population II stars
in the subgiant branch region of
the color-magnitude diagram has not changed all that much.
The results of this paper are limited by sampling statistics;
the RYI models are certainly good enough for the analysis of the
the small number of subgiants available in these observations.

The technique outlined in this paper would greatly benefit from
the use of stellar evolution models with fine age
and metallicity resolutions
(e.g. $\Delta t\!=\!0.1$ Gyr and $\Delta\!\feh=0.1$ dex, respectively).
Such fine age and metallicity resolutions
are not currently available with most published
theoretical isochrones.
If the availability of computation time or hardware constraints
are still important limitations,
then it would be most helpful if the authors of these theoretical models
would also provide accurate interpolation tools.
Timing information is very important in the study of star counts
in the subgiant branch region.
Young stars evolve more quickly through the subgiant branch stage
than do older (less massive) stars.
This leads to an incompleteness of the counts
of younger subgiant branch stars with respect to the older subgiant
branch stars.  This effect can be rectified if
timing information is available from evolutionary tracks.
Good observations are also necessary.
The derivation of stellar ages from the position of a star within the
subgiant branch region of the color-magnitude diagram assumes that
the luminosity and effective temperature of the star are not changing
rapidly.  Once stars evolve through the instability strip
(e.g. $\ga$1.5 $M_\odot$ at near-solar metallicities:
Gautschey \& Saio \cite{gasa1996}),
the proper interpretation of the age of such subgiant branch stars will
become problematical.  Users of the technique presented in this
paper must be sure that the stars they are analyzing have magnitudes
that do not vary by more than $\sim$$0.01$ mag.

The analysis of future observations of Carina would benefit
from an accurate determination of its reddening and metallicity.
This could, for example, be done using the SRM method
(Sarajedini \cite{sa1994}) on a shallow but complete survey of
red giant branch stars down to about one magnitude below the
the horizontal branch ($V\approx21.7$ mag).
Field contamination, especially in the outer
regions of the galaxy, can be measured
by observing nearby background fields of comparable area and depth.
Spectroscopic observations of Carina's red giants by the Gemini or VLT
telescopes will allow researchers to
determine the metallicity distribution function of the galaxy.
Detailed knowlege about the chemical evolution of the galaxy
would provide very useful observational constraints during the determination
of a detailed account of Carina's complex global star formation history.

\acknowledgments
We would like to thank
Taft Armandroff,
Ata Sarajedini,
Brian Chaboyer,
Doug Geisler,
Pierre Demarque,
Don VandenBerg,
Jeff Kuhn,
and
Gene Byrd
for their informative discussions and helpful comments
about various aspects of this project.
Support for this work was provided by a grant from
the National Aeronautics and Space Administration (NASA),
Order No.\ S-67046-F, which was awarded by
the Long-Term Space Astrophysics Program NRA 95-OSS-16.
This research has made use of
NASA's Astrophysics Data System Abstract Service
and the NASA/IPAC Extragalactic Database (NED)
which is operated by the Jet Propulsion Laboratory at Caltech, under
contract with the NASA.

\newpage

\appendix
\section{REVISING THE DISTANCE MODULUS OF CARINA}

Smecker-Hane \et (\cite{smet1994}) determined
the distance modulus of the Carina dwarf spheroidal galaxy to be
$(m-M)_0 = 20.12\!\pm\!0.08$
mag from
the mean $V$ magnitude of the horizontal branch,
$V_{\rm HB} = 22.65\!\pm\!0.05$ mag,
and the assumption that
the absorption in $V$ is
$A_V = 0.08$ mag
and that
$M_V^{\rm HB} = M_V^{\rm RR} = 0.45\!\pm\!0.07$ mag.
The estimate for
$M_V^{\rm HB}$ was derived
using an assumed metallicity for Carina of
$\feh \approx -2.2\!\pm\!0.4$ dex
with the
Lee \et (\cite{leet1990}, hereafter LDZ)
distance scale which gives the absolute $V$ magnitude for RR Lyrae
variables as
\begin{equation}
M_{V,{\rm{LDZ}}}^{\rm{RR}} = 0.17\feh + 0.82
\label{eq:LDZ}
\end{equation}
for a helium abundance on the main sequence $Y_{\rm{MS}}=0.23$.
Based on these assumptions, the $V$ apparent distance modulus
of Carina is $(m-M)_V^{~}= 20.20\!\pm\!0.08$ mag.

Lee \et (\cite{leet1993b}) have shown that
the $I$ magnitude of the tip of the red giant branch (TRGB)
of low-mass stars is a good distance indicator for resolved galaxies
with old metal-poor ($\feh<-0.7$ dex) stellar populations.
Observationally, the absolute $I$ magnitude of the tip of the red giant branch
(on the LDZ distance scale)
changes little for metallicities less than $-0.7$ dex :
$M_I^{\rm{TRGB}} \approx -4.0\!\pm\!0.1$ mag.
Smecker-Hane \et (\cite{smet1994})
derived a distance modulus for Carina of
$(m-M)_0^{~} = 20.05\!\pm\!0.09$ mag
using the TRGB method
(Da Costa \& Armandroff \cite{dacoar1990},
see also Lee \et \cite{leet1993b})
with their measurement of the apparent $I$ magnitude of the tip of the
red giant branch,
$I^{~}_{\rm{TRGB}} = 16.15\!\pm\!0.05$ mag,
and their estimate of the color of the TRGB,
$(B-I)_0^{\rm{TRGB}} = 2.85\!\pm\!0.05$ mag.
This derivation assumed that the metallicity of Carina
is $\feh=-2.0\!\pm\!0.4$ dex,
the absorption in $I$ is $A_I=0.05$ mag,
the reddening E$(B-I)$ is $0.05$ mag, and
$(B-V)/(V-I)=0.52\!\pm\!0.01$ mag for stars on the upper red giant branch.
Based on these assumptions, the $I$ apparent distance modulus
of Carina is $(m-M)_I^{~}= 20.10\!\pm\!0.09$ mag.

Smecker-Hane \et (\cite{smet1994})
derive a distance modulus of
$(m-M)_0^{~} = 20.09\!\pm\!0.06$ mag for Carina
from their measurements of $V_{\rm{HB}}$ and $I^{~}_{\rm{TRGB}}$.
This value is the average between
$20.12\!\pm\!0.08$ mag
and
$20.05\!\pm\!0.05$ mag
and the error of $0.06$ mag is
presumably derived from the simple propagation of errors which assumes
that the two determinations were uncorrelated.
The derivation of the $V$ apparent distance modulus is directly based
on the LDZ distance scale.
The derivation of the $I$ apparent distance modulus is based
the TRGB method which in turn is based on the work of
Da Costa \& Armandroff (\cite{dacoar1990})
who used the LDZ distance scale.
Since both methods use the LDZ distance scale,
the uncertainty of the zeropoint of the
$M_{V,{\rm{LDZ}}}^{RR}$ relation should not be ignored.
The distance modulus of Carina reported by Smecker-Hane \et (\cite{smet1994})
must have an error at least as large as our current uncertainty of
the error associated with the zeropoint of the $M_V^{\rm{RR}}$ relation:
$M_V^{RR} = -0.60\!\pm\!0.08$ mag at $\feh=-1.9$ dex
(Chaboyer \et \cite{chet1996a}).
If we conservatively add in quadrature
the $0.06$ mag internal errors with a
$0.08$ mag zeropoint error,
we can derive a reasonable estimate of the uncertainty of the
Smecker-Hane \et (\cite{smet1994})
distance modulus for Carina :
$(m-M)_0^{~} = 20.09\!\pm\!0.10$ mag.
Similarly, the
Smecker-Hane \et (\cite{smet1996})
distance modulus for Carina should presumably be revised to
$(m-M)_0^{~} = 20.12\!\pm\!0.09$ mag.

Da Costa and Hatzidimitriou determined that the
metallicity of the Carina dSph galaxy is
$\feh = -1.88\!\pm\!0.08$ dex
based on spectroscopic observations of the
\ion{Ca}{2} triplet
in Carina red giants
(Da Costa \cite{daco1993}).
Using a metallicity estimate of
$\feh = -1.9\!\pm\!0.2$ dex
for Carina
with the previous observations of
$V_{\rm{HB}}^{~}$ and $I_{\rm{TRGB}}^{~}$
by Smecker-Hane \et (\cite{smet1994}),
we have determined that the $V$ and $I$ apparent distance moduli
of Carina on the LDZ distance scale are, respectively,
$20.15\!\pm\!0.08$ mag
and
$20.08\!\pm\!0.09$ mag.

Observational evidence suggests that the zeropoint for the LDZ relation
[Equation (\ref{eq:LDZ})]
is about $0.10$ mag too bright at $\feh=-1.9$ dex.
Chaboyer \et (\cite{chet1996a},\cite{chet1996b}, hereafter CDKKS)
reviewed the literature and they advocate the use of the following relation
\begin{equation}
M_{V,{\rm{CDKKS}}}^{\rm{RR}} =
  { 0.20 \brack \pm0.04} \feh
+ { 0.98 \brack \pm0.08}
\label{eq:CDKKS}
\end{equation}
to determine the absolute $V$ magnitude of RR Lyrae stars.
This equation gives
$M_{V,{\rm{CDKKS}}}^{\rm{RR}} = 0.60\!\pm\!0.11$ mag
for $\feh=-1.9$ dex.
The absolute bolometric magnitude of RR Lyraes
on the CDKKS distance scale can be derived
by subtracting the $V$ bolometric correction for RR Lyraes that
was used by LDZ,
$BC_{V}^{RR} = -0.03\feh + 0.01\label{eq:bcvrr}$,
from Equation (\ref{eq:CDKKS}):
\begin{equation}
M_{\rm{bol,CDKKS}}^{\rm{RR}} =
  {0.23 \brack \pm0.04} \feh
+ {0.97 \brack \pm0.08}
\label{eq:CDKKSmbol}
{}~.
\end{equation}
The $V$ bolometric correction for a RR Lyrae star with
$\feh=-1.9$ dex is thus
$0.07$ mag which implies that the
absolute bolometric magnitude of the star is
$M_{V,{\rm{bol,CDKKS}}}^{\rm{RR}} = 0.53\!\pm\!0.11$ mag.
The difference between the LDZ and the CDKKS distance scales
at a metallicity of $\feh=-1.9$ dex is
$\Delta M_{\rm{CDKKS}}^{\rm{LDZ}} \approx 0.10$ mag
with $M_{V}^{\rm{RR}}$
on the CDKKS distance scale being fainter (larger).
Using a metallicity estimate of
$\feh = -1.9$ dex
for Carina
with the previous observations of
$V_{\rm{HB}}^{~}$ and $I_{\rm{TRGB}}^{~}$
by Smecker-Hane \et (\cite{smet1994}),
we have determined that the $V$ and $I$ apparent distance moduli
of Carina on the CDKKS distance scale are, respectively,
$20.05\!\pm\!0.11$ mag
and
$19.98\!\pm\!0.12$ mag.

Mould \& Aaronson (\cite{moaa1983})
used a reddening of $\ebmv=0.025\!\pm\!0.01$ mag
which was based on the reddening map of
Burstein \& Heiles (\cite{buhe1982})
and
Burstein (1983, private communication to Mould \& Aaronson).
Almost all subsequent studies of Carina have either used this reddening
or derived reddenings based on this value.
Now that a good estimate of the metallicity of Carina is available,
it is possible to directly
determine the reddening of Carina from CCD photometry.
Sarajedini (\cite{sa1994}) analyzed the observed red giant branches
of Da Costa \& Armandroff (\cite{dacoar1990}) and has determined that
the $\evmi$  reddening is
\begin{equation}
\evmi = (\vmi)_g - 0.1034\feh - 1.100
\label{eq:evmi}
\end{equation}
where $(\vmi)_g$ is the
the apparent $\vmi$ color of the red giant branch at the level
of the horizontal branch.
We have determined from the present observations that
the apparent $\vmi$ color of the red giant branch at the level of the
old horizontal branch is
$(\vmi)_g = 0.98\!\pm\!0.02$ mag.
With an assumed metallicity of $\feh=-1.9$ dex, we find that
the $\evmi$ reddening of Carina is $0.08\!\pm\!0.02$ mag.
The $\ebmv$ reddening is $0.06\!\pm\!0.02$ mag
assuming that
$\evmi \approx 1.3 \, \ebmv$
[Dean \et \cite{deet1978}].
This determination of $\ebmv$ is higher than but consistent with
(at the $\sim\,1.6$$\sigma$ level) the value
used by Mould \& Aaronson (\cite{moaa1983}).
The absorption in $V$ is determined to be
$A_V^{~}=0.19\!\pm\!0.06$ mag assuming that
$A_V^{~}=3.1\,\ebmv$ [Savage \& Mathis \cite{sama1979}].
The absorption in $I$ is determined to be
$A_I^{~} = 0.11\!\pm\!0.06$ mag from the relation
$A_I~{~} = A_V - \evmi$ mag.

We calculate the distance modulus of the Carina dwarf spheroidal galaxy
to be $(m-M)_0^{~}=19.87\!\pm\!0.11$ mag.
This distance modulus is on the CDKKS distance scale and
its value is the average between our estimates for
$(m-M)_V^{~}-A_V^{~}$
and
$(m-M)_I^{~}-A_I^{~}$.
The error estimate includes $0.08$ mag internal error and a
$0.08$ mag zeropoint error for the $M_V^{\rm{RR}}$ relation.
The main difference between this distance modulus and those reported
by Smecker-Hane \et (\cite{smet1994}, \cite{smet1996}) is the $0.10$ change
due to the use of the CDKKS distance scale
instead of the LDZ distance scale.
Our higher absorption values for $A_V^{~}$ and $A_I^{~}$ account for
the remaining difference.
Kuhn \et (\cite{kuet1996}) have recently made CCD observations
of Carina's RR Lryaes and they find
$V_{\rm{HB}}=20.71\!\pm\!0.01$ mag
in the central region of the galaxy and
$V_{\rm{HB}}=20.76\!\pm\!0.03$ mag
in the outer regions of Carina.
These new measurements of the $V$ magnitude of the horizontal branch
give distance moduli estimates of
$19.92\!\pm\!0.12$ mag
and
$19.97\!\pm\!0.12$ mag,
respectively, on the CDKKS distance scale assuming our standard values
$M_{V,{\rm{CDKKS}}}^{\rm{RR}} = 0.60\!\pm\!0.11$ mag
for $\feh=-1.9$ dex
with
$A_V^{~}=0.19\!\pm\!0.06$ mag.

\newpage

\figcaption[mighell.fig1.ps]{
\label{fig-1}
The $V$ vs $\vmi$ color-magnitude diagram and apparent $V$ stellar luminosity
function of the observed stellar field in the Carina dwarf spheroidal galaxy.
The error bars in the color-magnitude diagram indicate r.m.s. (1 $\sigma$)
uncertainties for a single star at the corresponding magnitude.
The error bars for the apparent $V$ stellar luminosity function
are derived from counting (Poisson) statistics.
This figure shows 2025 stars down to $V\,=\,26.0$
mag and 3,185 stars down to a limiting magnitude of $V\,\approx\,27.2$ mag
on the main sequence of Carina.
The {\em Hubble Space Telescope} WFPC2 instrument was used to make two 2200 s
observations with the F555W and F814W filters.
}

\figcaption[mighell.fig2.ps]{
\label{fig-2}
The apparent $V$ stellar luminosity function of the Carina WFPC2 field is
shown (filled circles) with the Carina stellar luminosity function
(open diamonds) of Mighell (\protect\cite{mi1990c}) which has been
scaled down by a factor of 3.2.
The open diamonds are plotted 0.04 mag too faint
to improve readability.
For comparision, the 6, 8, and 14 Gyr
theoretical RYI (Green \et \protect\cite{gret1987})
stellar luminosity functions for metallicity
$Z=0.00025$ ($\feh = -1.9$) and helium abundance $Y=0.24$
are respectively displayed as the solid, dashed, and dotted curves.
The RYI stellar luminosity functions are plotted
using an apparent $V$ distance modulus of $(m-M)_V=20.05$ mag
and have been normalized to have the same number of stars as the apparent
Carina stellar luminosity function in the magnitude range
$25.0\!\leq\!V\!\leq\!26.0$.  See the text for further details.
}

\figcaption[mighell.fig3.ps]{
\label{fig-3}
The $V$ vs $\vmi$ color-magnitude diagram of the main-sequence turnoff region
of the Carina dwarf spheroidal galaxy. The error bars indicate $1\!\sigma$
uncertainties for a single star at the corresponding magnitude. The
theoretical RYI isochrones for ages
2 to 18 Gyr (left to right) are displayed with 2 Gyr steps for $Z=0.00025$
($\feh = -1.9$)
and $Y=0.24$. The isochrones have been displayed using an apparent $V$
distance modulus $(m-M)_V=20.05$ mag and a reddening $\evmi=0.08$ mag.
}

\figcaption[mighell.fig4.ps]{
\label{fig-4}
The $V$ vs $\vmi$ color-magnitude diagram of the main-sequence turnoff region
of the Carina dwarf spheroidal galaxy. The error bars indicate $1\!\sigma$
uncertainties for a single star at the corresponding magnitude. The
theoretical RYI isochrones for ages
2 to 18 Gyr (left to right, bottom to top) are displayed with 1 Gyr steps for
$Z=0.00025$ and $Y=0.24$. The isochrones have been displayed using an
apparent $V$ distance modulus $(m-M)_V=20.05$ mag and a reddening $\evmi=0.08$
mag. The interpolated isochrones have also been plotted (at 0.25 Gyr
increments) from a point on the isochrone near
the main-sequence turnoff (long-dashed curve) to the base of the
red giant branch [short-dashed line at $(\vmi)_0=0.70$ mag]. See the text for
further details.
}

\figcaption[mighell.fig5.ps]{
\label{fig-5}
This figure shows the 1000 simulated observations (dots)
for all of the 152 stars (asterisks)
found in the bounded region of Fig.\ 4.
The scatter of the simulated observations for
an individual star appears to be
elliptical principally due to the standard 5:1 plotting ratio of the
magnitude scales ($V$:$\vmi$) used in the figure.
A total of 92.7\% of the simulated observations were
found within the bounded region.
The other details of the figure are described in Fig.\ 4.
See the text for further details.
}

\figcaption[mighell.fig6.ps]{
\label{fig-6}
The r.m.s.\ errors of the stellar age estimates increase with age
because photometric errors increase with apparent magnitude
and the spacing between theoretical isochrones generally decrease
with increasing age.
This figure shows the distribution of estimated ages for the
1000 simulated observations of the stars
246342968 (dashed histogram),
409901113 (dotted histogram), and
269902650 (solid histogram),
which have an estimated stellar ages of
$5.04\!\pm\!0.25$,
$6.87\!\pm\!0.35$,
and
$9.70\!\pm\!0.52$
Gyr, respectively.
The figure shows that the simulated observations
of these stars were spread, respectively,
over 2, 6, and 10 bins that are 0.25 Gyr wide.
}

\figcaption[mighell.fig7.ps]{
\label{fig-7}
The mean estimated ages (in Gyr on the RYI age scale)
and r.m.s.\ errors
of all 152 stars in the bounded region of Fig.\ 4.
Table 3 is the source of the data.
There are 116 stars (filled squares) that had at least 897 of the 1000
simulated observations found within the bounded region of Fig.\ 4
($p\geq0.897$).
The 36 remaining stars are shown as open squares.
See the text for further details.
}

\figcaption[mighell.fig8.ps]{
\label{fig-8}
This figure shows the 1000 simulated observations (dots)
for the 116 stars (asterisks)
found in the bounded region of Fig.\ 4 with at least a 99\% probability
of actually being found within the bounded region.
The remaining 36 stars (squares)
out of the original 152 found bounded region are also shown.
The other details of the figure are as described in Fig.\ 5.
See the text for further details.
}

\figcaption[mighell.fig9.ps]{
\label{fig-9}
The estimated mean stellar ages
(in Gyr on the RYI age scale)
of the conservative subset of 116 subgiant branch stars
($p\!\geq\!0.897 \Rightarrow p^\prime\!\geq\!0.990$).
The errors for all these stars are shown as filled squares in Fig.\ 7.
Two 0.25 Gyr wide bins were added together
to produce 2 low-resolution star-count histograms with 0.5 Gyr bin widths
that are offset by one bin from each other:
the vertical-shaded histogram
gives the counts at $2.00, 2.50, 3.00, 3.50, \ldots$ Gyr;
the horizontal-shaded histogram
gives the counts at $2.25, 2.75, 3.25, 3.75, \ldots$ Gyr.
See the text for further details.
}

\begin{table}
\dummytable\label{tbl-1}
\end{table}

\begin{table}
\dummytable\label{tbl-2}
\end{table}

\begin{table}
\dummytable\label{tbl-3}
\end{table}

\begin{table}
\dummytable\label{tbl-4}
\end{table}

\begin{table}
\dummytable\label{tbl-x}
\end{table}

\end{document}